\documentclass[%
 reprint,
superscriptaddress,
 amsmath,amssymb,
 aps,
 pra,
]{revtex4-1}

\usepackage{color}
\usepackage{graphicx}
\usepackage{bm}
\usepackage{siunitx}
\usepackage{hyperref}

\renewcommand{\Re}{\operatorname{Re}}
\renewcommand{\Im}{\operatorname{Im}}

\begin{document}

\title{Characterizing twin-particle entanglement in double-well potentials}

\author{Marie Bonneau}
\email{marie.bonneau@tuwien.ac.at}
\affiliation{ Vienna Center for Quantum Science and Technology, Atominstitut, TU Wien, Stadionallee 2, 1020 Vienna, Austria}

\author{William J. Munro} 

\affiliation{NTT  Basic  Research  Laboratories,  3-1  Morinosato-Wakamiya, Atsugi,  Kanagawa  243-0198,  Japan}
\affiliation{National  Institute  of  Informatics,  2-1-2  Hitotsubashi, Chiyoda-ku,  Tokyo  101-8430,  Japan}

\author{Kae Nemoto}
\affiliation{National  Institute  of  Informatics,  2-1-2  Hitotsubashi, Chiyoda-ku,  Tokyo  101-8430,  Japan}

\author{J{\"o}rg Schmiedmayer}

\affiliation{ Vienna Center for Quantum Science and Technology, Atominstitut, TU Wien, Stadionallee 2, 1020 Vienna, Austria}


\begin{abstract}

We consider a pair of twin atoms trapped in double-well potentials. For each atom, two orthogonal spatial modes are accessible: the states $ |L\rangle$ and $|R\rangle$ spatially localized in the left and right wells respectively. Furthermore the twin atoms are distinguishable thanks to an additional degree of freedom. We propose a method for experimentally quantifying the particle entanglement between these atoms which allows us to probe a violation of Bell's inequality. It is based on measuring the correlations in the atoms' momentum distribution. 
If the tunneling and the energy difference between the wells are tunable, then full state tomography is achievable.
\end{abstract}



\maketitle

\section{Introduction}
\label{sec:intro}

A key resource necessary for the development of quantum technologies is the availability of entangled-state sources. For example in quantum optics a widely used source of entangled photons is parametric down-conversion, where twin photons are emitted in correlated spatial modes \cite{Burnham1970,Walborn2010}. In the atomic realm, twin-atom sources have been developed based on atomic four-wave mixing in Bose-Einstein condensates (BEC) \cite{Perrin2007,Bucker2011,WuRuGway2011,Bookjans2011,Lucke2011,Gross2011,Bonneau2013}. The two output modes of these four-wave mixing processes are non-degenerate such that the twin atoms have distinct momentum or spin states. 
Over the last few years twin atoms have been demonstrated to be momentum-correlated \cite{Perrin2007, Bucker2011, Bonneau2013} and EPR entangled \cite{Peise2015a}. 
Twin-atom sources have been applied to atom interferometry beyond the standard quantum limit \cite{Lucke2011, Kruse2016}, to observation of the Hong-Ou-Mandel effect with atoms in free space \cite{Lopes2015} and to generation of entanglement between different spatial regions \cite{Fadel2018,Kunkel2018,Lange2018a}. 
The next challenge is to run these sources in configurations where single pairs of twin atoms can be used as pairs of entangled qubits.\\

\begin{figure*}
\centering
\def\svgwidth{\columnwidth}
\includegraphics[width=\textwidth]{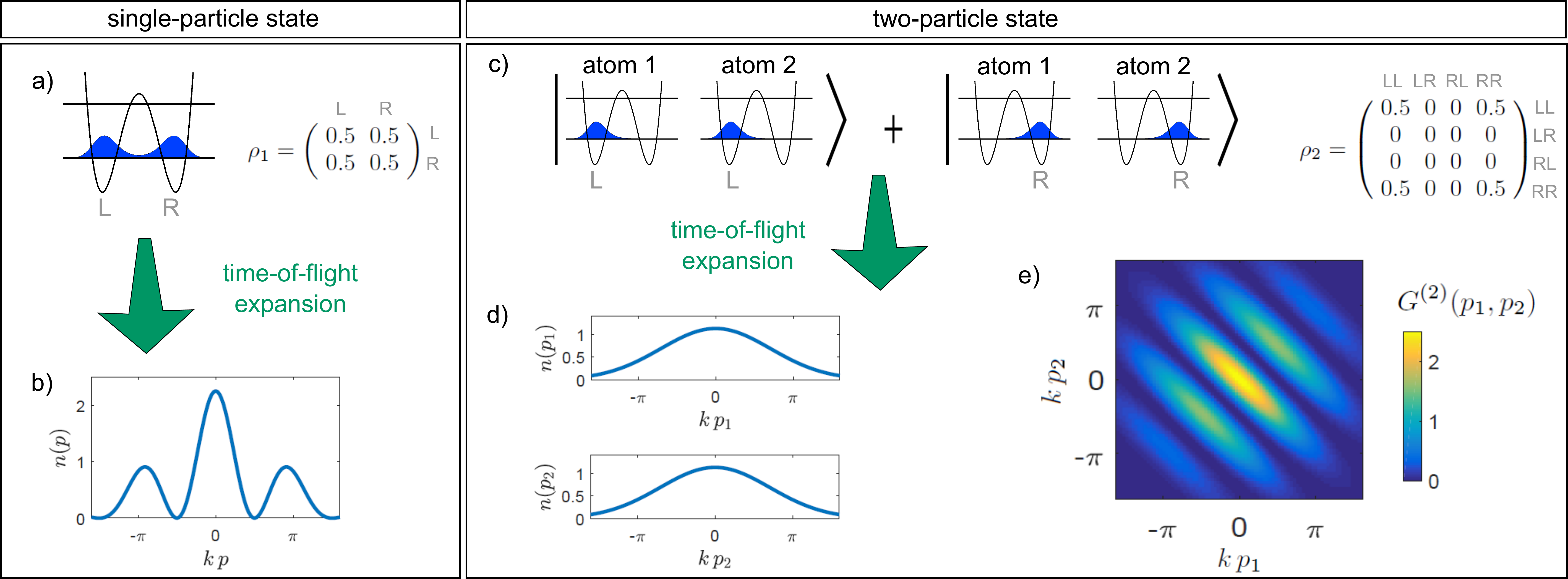}
\caption{\label{fig:dw} Measuring the momentum distribution of atoms trapped in a double-well potential is similar to measuring the far-field intensity distribution in a double-slit experiment. For atoms in the single-particle pure state $\rho_1$ (left), interferences in the momentum distribution $n(p)$ (b) reflect the coherences of the density matrix and the position distribution (a) the populations. For atoms in a non-separable two-particle state (c), the mean momentum distributions $n(p_1)$, $n(p_2)$ of each particle separately (d) are not enough to characterize the state, while interferences in the momentum second-order correlation (e) relate to the coherences of the two-particle density matrix.}
\end{figure*}

An atom from a twin-atom source can be trapped in a double-well potential where the presence of that atom either in the left $|L\rangle$ or right $|R\rangle$ well forms the basis states of a qubit. These spatially localized modes (see Fig.~\ref{fig:dw}) provide a natural way to describe our twin-atom source and the entangled qubits it generates. In the experiments \cite{Perrin2007, Bucker2011, Bonneau2013, Lopes2015}, the twin atoms are emitted in two different momentum states which we will label 1 and 2. This momentum degree of freedom allows us to distinguish the two atoms emitted from our twin source and so our system in this case is composed of two distinguishable qubits, whose interactions are negligible. Alternatively, the $\{1, 2\}$ degree of freedom can correspond to two different spin states, like in experiments \cite{Bookjans2011,Lucke2011,Gross2011,Peise2015a,Kruse2016,Fadel2018,Kunkel2018,Lange2018a}. The description presented above is definitely not unique. We can also describe our source from an alternative point of view where we use both the position ($|L\rangle$, $|R\rangle$) and momentum degrees of freedom to define the atomic modes. In this view point, four single-atom modes in total are available for two indistinguishable atoms which we can represent as $|L\rangle_1$, $|R\rangle_1$, $|L\rangle_2$, and $|R\rangle_2$. Now for a single atom pair emitted from a twin-particle source, there is exactly one atom in state 1 and one atom in state 2. The two-atom state is therefore restricted to the symmetrized Hilbert space $ \mathcal{H}_{pair}= \{ |L,L\rangle , |L,R\rangle , |R,L\rangle , |R,R\rangle \}$, with  $|i,j\rangle$ defined by $|i,j\rangle= 1/\sqrt{2} \left( |i\rangle_1 \otimes |j\rangle_2  + |j\rangle_2 \otimes |i\rangle_1  \right)$. For experiments where single twin-atom pairs are emitted in an entangled two-atom state, it is important to investigate the correlations present within this state.\\

We present here an experimentally realizable protocol for performing full tomography of twin-atom states in double-well potentials. It applies to any pure or mixed state of $\mathcal{H}_{pair}$. The first part of this protocol does not require any manipulation in the $\{L, R\}$ space, thanks to correlation measurements in momentum space, but already allows testing whether the twin-atom state violates a Bell’s inequality in the $\{L, R\}$ degree of freedom.  The second part involves rotations of the  {L, R} space allowing all the remaining coherence elements of the quantum state to be determined and hence the full state tomography.

Furthermore our approach is also relevant to other two-atoms experiments in double-well potentials \cite{Kaufman2015, Murmann2015, Dai2016, Lester2018}, where exactly two atoms are selected and entanglement between the spin ($\{1,2\}$) and spatial ($\{L,R\}$) degrees of freedom is engineered. Recently, based on the hypothesis that there is exactly one atom in each well, this entanglement was characterized by performing rotations in spin space combined with \emph{in situ} population correlation measurements \cite{Dai2016, Lester2018}, and a Bell's inequality violation in spin space was demonstrated \cite{Dai2016}. In contrast, in our approach we know that there is exactly one atom in each of the $\{|1\rangle, |2\rangle \}$ states due to the twin character but do not need to know the initial configuration of the $\{L, R\}$ spatial modes. Our method could however be extended to situations with two indistinguishable atoms in a double-well potential \cite{Kaufman2014,Islam2015}.

The situation we consider here, i.e., two atoms in double-well potentials, is an analog to a double-slit experiment with two photons. The intensity distributions in the near-field and far-field play the role of the atom density distributions in position and momentum space. Such an experiment was first performed 30 years ago by Ghosh and Mandel \cite{Ghosh1987}: They sent twin photons, entangled in momentum, through a double slit. By measuring the two-photon coincidence detections in the far field they observed conditional interference fringes, which were shown to be related to entanglement \cite{Greenberger1993}. 
More recently several quantum optics experiments exploited interference patterns in coincidences for characterizing the state of two photons behind double slits, using different approaches: 
In Refs.~\cite{Neves2007,Peeters2009}, the two photons were assumed to be in a pure state and in a specific subspace of $ \mathcal{H}_{pair}$. Under these hypotheses, measuring the coincidences in the far-field on the one hand and the coincidences in the near-field on the other hand allows identifying the state and quantifying directly the entanglement from the contrast of the coincidence pattern. 
In ref.~\cite{Taguchi2008}, additionally, the correlations between the near and the far fields (or in some intermediate plane) could be measured, and the two-particle density matrix was fully reconstructed for various states of $ \mathcal{H}_{pair}$, without any assumption on these states. Finally, another method for full tomography of any state of $ \mathcal{H}_{pair}$ was demonstrated in ref.~\cite{Pimenta2010}, where a spatial light modulator was introduced in the near-field of the slits in order to rotate the measurement basis and the coincidences were measured independently in the near and far fields.\\

The method of Refs.~\cite{Neves2007,Peeters2009} applies to any system analogous to a double-slit where the near-field and far-field distributions are measured independently. An example are twin atoms in a double-well potential; another example is untrapped twin atom beams where two spatial modes are selected from each beam at the early stage of their propagation, using either Bragg selection \cite{Dussarrat2017} or a material mask \cite{Khakimov2016}, and overlap after time-of-flight (assuming the initial size of the spatial modes to be negligible with respect to their size after propagation). It was thus proposed to demonstrate the entanglement of twin atom beams emitted into given maximally entangled states from the interferences in coincidences, for example using squeezing \cite{Gneiting2011} or ghost interferences \cite{Kofler2012}.\\

We will thoroughly describe how the coincidence patterns in position and in momentum relate to the two-particle density matrix, such that for any state of $ \mathcal{H}_{pair}$ these two observable give access to most of the density matrix elements. Knowing these elements is sufficient for probing a violation of Bell's inequality in the $\{ |L\rangle , |R\rangle \}$ basis. Additionally, we propose a tomography method specific to the double-well configuration. It requires us to accurately manipulate the atomic external state by tuning the shape of the double-well potential \cite{Berrada2013}. Most cold-atom setups with double-well trapping potentials allow tuning the tunnel coupling between the two wells from a negligible value to a regime where Josephson oscillations occur \cite{Albiez2005}. In this regime, the information encoded on the relative phase of the two wells can be transferred to their population and vice versa. On top of this, tuning the energy difference between the two wells provides control of their relative phase. By combining these two operations before the detection of the atomic momentum distribution one can reconstruct all elements of the two-atom density matrix. With this method, performing a full tomography of two-atom states is thus remarkably simple.\\

Our article is organized as follow: In Sec.~\ref{sec:G2} we show how the correlations of the atomic position and momentum distributions relate to the density matrix. This leads to Sec.~\ref{sec:quant} where these correlations are used to set a lower bound on the system's entanglement and to Sec.~\ref{sec:Bell} where they are used to probe a violation of Bell's inequality. In Sec.~\ref{sec:tom} we explain how tunability of the trapping potential allows full tomography of any state of $ \mathcal{H}_{pair}$. Finally, we describe in Sec.~\ref{sec:xp} an experimental situation for which this method could be implemented.

\section{Information contained in two-atom correlations}
\label{sec:G2}

Our characterization of two-atom states is inspired by the reconstruction of single-atom states in a double-well potential \cite{Albiez2005,Berrada2013}. In the latter situation, the single-atom density matrix is expressed in the $\{|L\rangle, |R\rangle \}$ basis. The populations are determined from the mean \emph{in situ} density in each of the two wells, and the coherences from the mean density in momentum space. If the atoms are in a coherent superposition of the $|L\rangle$ and $|R\rangle$ states (see example in Fig.~\ref{fig:dw}.a), their mean momentum distribution exhibits fringes. The fringe contrast determines the amplitude of the density matrix coherences (or at least gives a lower bound for it) and the fringe position their phase. \\

Considering now a two-atom state, its density matrix $\hat{\rho}$ can only be characterized when measuring two-particle properties. Since the mean density is a one-particle property, measuring it will not be sufficient for characterizing the state. For example, for a fully entangled state (see Fig.~\ref{fig:dw}.b), no interference fringes are visible on the mean momentum distribution of each particle separately, since the reduced one-particle density matrix is in a totally mixed state. Therefore, in the following, we will look at the second-order correlation function 
\begin{equation}
\begin{split}
G^{(2)}(\xi_1,\xi_2) &= \langle \hat{\Psi^{\dag}}(\xi_1) \hat{\Psi^{\dag}}(\xi_2)  \hat{\Psi}(\xi_1) \hat{\Psi}(\xi_2)\rangle \\
                       &= \operatorname{Tr} [  \hat{\rho} \; \hat{\Psi^{\dag}}(\xi_1) \hat{\Psi^{\dag}}(\xi_2)  \hat{\Psi}(\xi_1) \hat{\Psi}(\xi_2) ]
\end{split}
\end{equation}
 where $\xi_1$ ($\xi_2$) is the coordinate of the first (second) atoms while the field $\hat{\Psi}(\xi_n)$  describes the $n$-th atom and can be expressed in the $\{|L\rangle, |R\rangle \}$ basis as  
\begin{equation}
{\hat{\Psi}(\xi_n) = \sum_{i=L,R} \Phi_i(\xi_n) \; \hat{a}_{ni} }
\end{equation}
   with $\hat{a}_{ni}$ being the annihilation operator of the $n$-th atom in the mode $|i\rangle$ of wave function $\Phi_i$. Decomposing the two-atom density matrix as
\begin{equation}
\hat{\rho}=  \sum_{i,j,k,l=L,R} \rho_{ijkl} \; |i,j\rangle \langle k,l| 
\end{equation} 
and rewriting the two-atom states $|i,j\rangle$ in second quantization (with $n_{l _{1/2 }}$ the occupation of the $| l \rangle _{1/2 }$ mode, $l =L,R $):
\begin{equation}
|n_{L_1}=\delta_{i,L} \, ; \, n_{R_1}=\delta_{i,R} \, ; \, n_{L_2}=\delta_{j,L} \, ; \, n_{R_2}=\delta_{j,R}\rangle
\end{equation}
one obtains the following relation between the second-order correlation function and the elements of the density matrix:
\begin{equation}
\label{eq:G2G}
\begin{split}
G^{(2)}(\xi_1,\xi_2)= \sum_{i,j,k,l=L,R} \rho_{ijkl} \; \Phi_i(\xi_1) \Phi_j(\xi_2) \Phi_k^*(\xi_1) \Phi_l^*(\xi_2) 
\end{split}
\end{equation}
Now $G^{(2)}(\xi_1,\xi_2)$ is experimentally accessible from the density distributions $n(\xi_1)$, $n(\xi_2)$ of the first and second atoms: 
\begin{equation}
G^{(2)}(\xi_1,\xi_2)= \langle n(\xi_1) \,  n(\xi_2)\rangle
\end{equation}
Importantly, this requires identifying the partner of each atom, which means detecting single pairs. The coordinates $\xi_1$ and $\xi_2$ refer either to the position or to the momentum of the atoms, in both cases along the double-well axis. In analogy to the single-particle situation, the populations of $\hat{\rho}$ will be extracted from measurements of $G^{(2)}$ in position space and the coherences from interference patterns arising in $G^{(2)}$ in momentum space.\\


\subsection{Correlations in position: Populations}

In position space, the $|L\rangle$ and $|R\rangle$ modes are spatially separated \footnote{Or can be mapped on spatially separated modes, as used for example in ref.~\cite{Berrada2013}.}. Therefore, Eq.~(\ref{eq:G2G}) simplifies to 
\begin{equation}
G^{(2)}(x_1,x_2)= \sum_{i,j=L,R} \rho_{ijij} \; |\Phi_i(x_1)|^2 |\Phi_j(x_2)|^2
\end{equation}
When integrating over the region $D_L$ ($D_R$) containing the $|L\rangle$ ($|R\rangle$) mode, one obtains the populations of the two-atom density matrix as
\begin{equation}
\begin{split}
p_{ij} &= \rho_{ijij} = \int_{D_i} dx_1 \int_{D_j} dx_2 \;  G^{(2)}(x_1,x_2)\\
       &=   \int_{D_i} dx_1   \int_{D_j} dx_2 \; \left \langle   n(x_1) \; n(x_2) \right \rangle\\
       &= \langle n_{1i} \;  n_{2j} \rangle 
\end{split}
\end{equation}
where $n_{1L}$, $n_{1R}$, $n_{2L}$ and $n_{2R}$ are the occupations of the left and right well by the first and second atom.


\subsection{Correlations in momentum: Coherences}

\begin{figure}
\centering
\def\svgwidth{\columnwidth}
\includegraphics[width=\columnwidth]{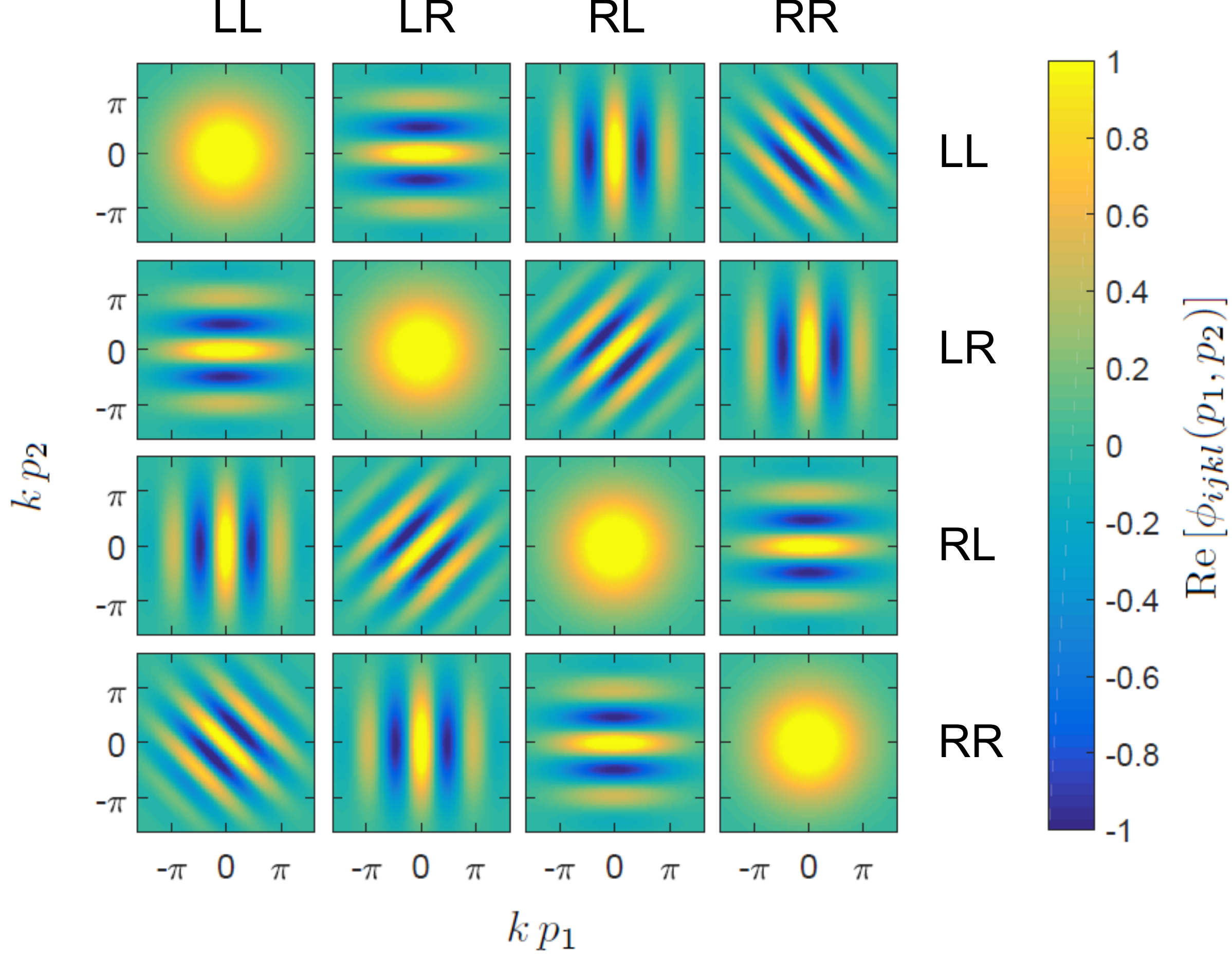}
\caption{\label{fig:G2d} Different terms $\rho_{ijkl}$ of a two-qubit density matrix $\rho$ contribute to the pattern of $G^{(2)} (p_+,p_-)$ with different patterns $\phi_{ijkl}(p_+,p_-)$, whose real part is plotted here.}
\end{figure}

\begin{figure*}
\centering
\def\svgwidth{\columnwidth}
\includegraphics[width=\textwidth]{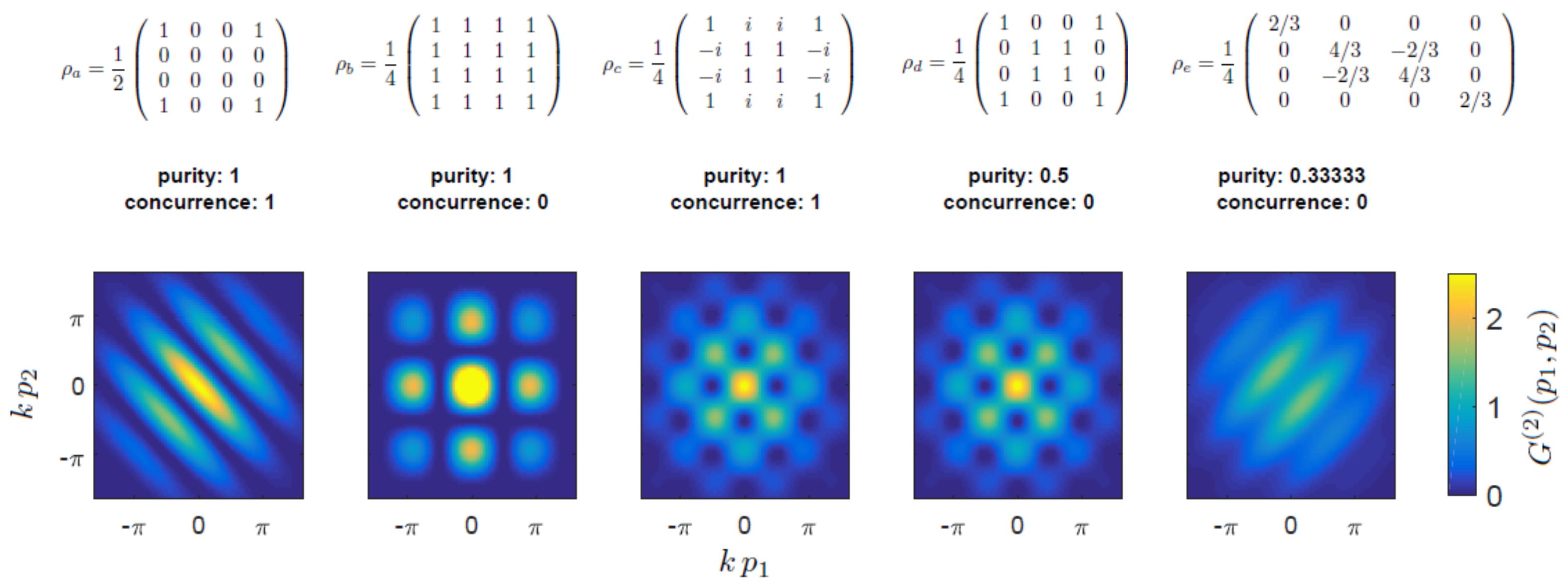}
\caption{\label{fig:G2e} Examples of correlation functions in momentum space $G^{(2)} (p_1,p_2)$ (bottom row) for two-atom states (top row) of different purities and concurrences (middle row). Spatially non-separable patterns can be obtained not only for entangled pure states like  $\rho_a$ and  $\rho_c$, but also for mixed states like $\rho_d$ and the Werner state $\rho_e$. For the product state $\rho_b$, $G^{(2)} (p_1,p_2)$ is simply the product of the oscillating mean densities of the two atoms. $G^{(2)} (p_1,p_2)$ provides information on some terms only of the two-atom density matrix $\rho$ (see text). Therefore the fully entangled state $\rho_c$ and the fully mixed state $\rho_d$ exhibit the same checkerboard pattern. However, in the particular case of the state $\rho_a$, the maximal contrast of the fringe pattern allows full reconstruction of the state.} 
\end{figure*}

Considering now the momentum space, we can rewrite Eq.~(\ref{eq:G2G}) as
\begin{equation}
\label{eq:G2comp}
G^{(2)} (p_1,p_2) = \sum_{i,j,k,l \in \{L,R\} } \phi_{ijkl}(p_1,p_2) \; \rho_{ijkl}
\end{equation}
where
\begin{equation}
\label{feq}
\phi_{ijkl}(p_1,p_2) =\Phi_i(p_1)  \Phi_j(p_2) \Phi_k^*(p_1) \Phi_l^*(p_2)
\end{equation}
The momentum-space wave functions of the $|L\rangle$ and $|R\rangle$ modes are $\Phi_L (p) =  e^{- i k p}  \mathcal{A}(p) $ and $\Phi_R (p) =  e^{ i k p}   \mathcal{A}(p) $ respectively, where $k$ is determined by the wells spacing. The exact shape of the envelop $\mathcal{A}(p)$ is not relevant for this discussion. For simplicity we use in Fig.~\ref{fig:G2d} a Gaussian $\mathcal{A}(p)= e^{- p^2/2 \sigma_p^2} ( 2 /\sqrt{\pi} \sigma_p)$, which is a good approximation for no tunnel coupling between the wells only. $\sigma_p$ is defined by the width of the $|L\rangle$ and $|R\rangle$ wave functions in position space and is set to $0.5/k$ in Figs.~\ref{fig:G2d} and \ref{fig:G2e}.\\

Let us now focus on the spatial dependence of the $\phi_{ijkl}(p_1,p_2)$ functions. From Eq.~(\ref{feq}), one gets
$$\phi_{ijkl}(p_1,p_2)=\mathcal{A}(p_1)^2 \mathcal{A}(p_2)^2  F_{ijkl}(p_1,p_2),$$
where $F_{ijkl}(p_1,p_2)$ is given by\\
\\
\begin{tabular}{|c|c|c|c|c|}
\hline
$ij \backslash kl$ & $LL$ & $LR$ & $RL$ & $RR$ \\
\hline
$LL$ & 1 & $e^{-2 i k p_2 }$ & $e^{-2 ik  p_1}$ & $e^{-2 i k (p_1+p_2) }$ \\
\hline
 $LR$ & $e^{2 i k p_2 }$ & 1 & $e^{-2 i k (p_1 - p_2) }$ & $e^{-2 i k p_1 }$ \\
\hline
$RL$ & $e^{2 i k p_1 }$ & $e^{2 i k (p_1-p_2) }$ & 1 & $e^{-2 i k p_2 }$ \\
\hline
$RR$ & $e^{2 i k (p_1 + p_2) }$ & $e^{2 i k p_1 }$ & $e^{2 i k p_2 }$ & 1\\
\hline
\end{tabular}\\
\\
In Fig.~\ref{fig:G2d} we plot the real part of the $\phi_{ijkl}(p_1,p_2)$ functions. No fringe pattern is visible for the contributions of the populations to $G^{(2)} (p_1,p_2)$, while the contributions of the coherence terms exhibit oscillations along one axis when they involve both the $|L\rangle$ and the $|R\rangle$ modes of the corresponding atom. Therefore the contributions of the anti-diagonal terms ($\phi_{LRRL}$ and $\phi_{LLRR}$) are diagonal and antidiagonal fringes. The other coherence terms are horizontal (vertical) fringes, each term being identical to a second one: $\phi_{LLRL}=\phi_{LRRR}$ and $\phi_{LLLR}=\phi_{RLRR}$. For a given state $\rho$, $G^{(2)} (p_1,p_2)$ is obtained by combining these contributions following Eq.~(\ref{eq:G2comp}), as shown for example on Fig.~\ref{fig:G2e}. Only the real part of the $\phi_{ijkl}(p_1,p_2)$ functions contribute to $G^{(2)} (p_1,p_2)$.\\


Let us now assume that we want to characterize an unknown state of $\mathcal{H}_{LR}$. We have already described how the populations can be extracted from the correlations in position space. We now would like to extract information on the coherences from the correlation function in momentum space. 
For this, we need to determine the weights of the 4 independent $\phi_{ijkl}(p_1,p_2)$ fringe patterns in the measured $G^{(2)}(p_1,p_2)$. In the following we call these weights 
$a=\rho_{LRRL}$, $b=\rho_{LLRR}$, $y=\rho_{LLRL}+\rho_{LRRR}$ and $z=\rho_{LLLR}+\rho_{RLRR}$. 
These are complex values, used as free parameters when fitting $G^{(2)}(p_1,p_2)$ by the expression

\begin{widetext}
\begin{equation}
g(a,b,c,d,p_1,p_2) = \phi_{LLLL} + 2 \Re(a  \;  \phi_{LRRL}) + 2 \Re(b  \; \phi_{LLRR}) + 2 \Re(y \; \phi_{LLRL}) + 2 \Re(z  \; \phi_{LLLR})
\end{equation}
\end{widetext}

Their module is given by the amplitude of the fringes and their phase is given by the fringe position. The parameters $a$ and $b$ thus give us directly two of the coherence terms.\\ 

As a summary, measuring two-particle correlations in position and momentum space provides the following information on the two-atom density matrix in the $\{|L,L\rangle, |L,R\rangle, |R,L\rangle, |R,R\rangle \}$ basis (elements in bold):
\begin{equation}
\label{eq:rho}
\rho= \left( \begin{array}{cccc}
\mathbf{p_{LL}} & c &  d & \mathbf{b} \\
c* & \mathbf{p_{LR}} & \mathbf{a} & e \\
d* & \mathbf{a*} & \mathbf{p_{RL}} & f \\
\mathbf{b*} & e* &  f* & \mathbf{p_{RR}}  \end{array} \right).
\end{equation}
The complex coefficients $c$, $d$, $e$, and $f$ are still unknown but are linked by the constraints $c+f=z$ and  $d+e=y$, where $z$ and $y$ are obtained from the fit. Furthermore, since $\rho$ is a density matrix, its eigenvalues should be real and belong to $[0,1]$. This  allows setting some bounds on the $c$, $d$, $e$, and $f$ coefficients. In particular, for the Bell state $\rho_a$ of Fig.~\ref{fig:G2e} (as well as for other Bell states), the above-described measurements allow reconstructing the full density matrix without making any assumption on the state (the remaining coherence terms being then 0).


\begin{figure*}
\centering
\def\svgwidth{\columnwidth}
\includegraphics[width=\textwidth]{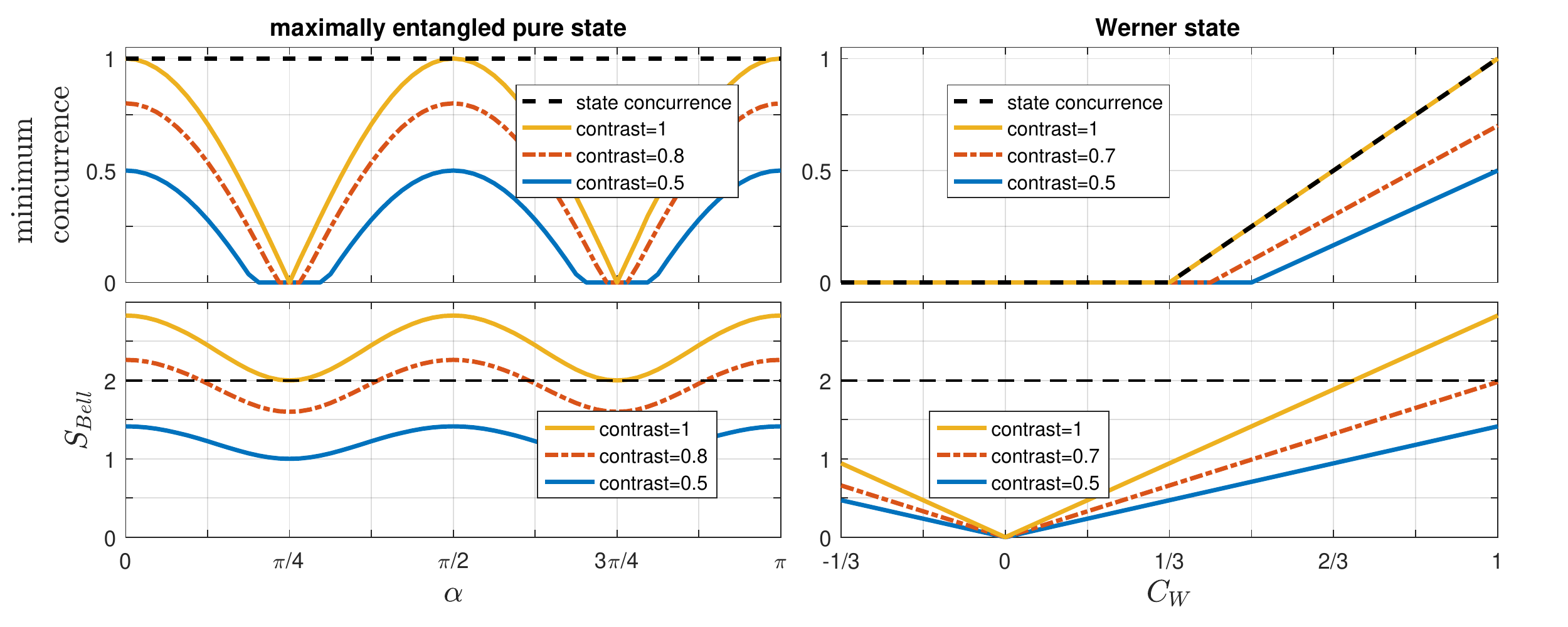}
    \caption{\label{fig:conc} Left: We consider here a maximally entangled state of the form $ |\Psi (\alpha)\rangle$ defined in Eq.~(\ref{eq:state}). Without assuming initial knowledge on this state, the experimentally measurable correlations allow setting a lower bound on its concurrence (top), characterizing its entanglement, and computing the parameter $S_{Bell}$ which describes a Bell inequality violation in the $\{|L\rangle, |R\rangle \}$ basis (bottom). 
Without experimental limitation on the fringes contrast (yellow/light gray line) one can in principle measure a maximal entanglement (the concurrence being 1) and a maximal violation of a Bell inequality ($S_{Bell}=2 \sqrt{2}$) for the Bell states $ |\Psi (0)\rangle = 1/\sqrt{2} \left( |LL\rangle +  |RR\rangle \right)$ and $ |\Psi (\pi /2)\rangle = 1/\sqrt{2} \left( |LR\rangle +  |RL\rangle \right)$. The effect of a finite experimental contrast reduces the lower bound on the concurrence and the region of the Bell violation [dash-dotted red and (lower) blue lines].
Right: In the case of a Werner state $\rho_W$ [defined in Eq.~(\ref{eq:Werner})], the correlation measurement in the $\{|L\rangle, |R\rangle \}$ basis captures well the state entanglement: the lower bound on the concurrence for a perfect contrast coincides with the concurrence itself (top, yellow/light gray and dashed black lines).}
\end{figure*}

\section{Entanglement estimation}
\label{sec:quant}

Although we have not yet determined the full density matrix, we can already put bounds on the entanglement of the state $\rho$. We quantify the entanglement with the concurrence defined as \cite{Wootters1998a}: 
\begin{equation}
C(\rho) =\text{max} (0, \lambda_1-\lambda_2-\lambda_3 - \lambda_4)
\end{equation}
with $\lambda_1 \geq \lambda_2 \geq \lambda_3 \geq \lambda_4$ the square roots of the eigenvalues of the matrix $\rho \, \left ( \sigma_2^1 \otimes \sigma_2^2  \right ) \, \rho^*  \,  \left ( \sigma_2^1 \otimes \sigma_2^2 \right )$ where the matrix $\sigma_2^j = \left( \begin{array}{cc}
0 & -i \\
i & 0 \end{array} \right)$ applies on the $j$-th atom. The concurrence is 1 for a maximally entangled state, and 0 for a non-entangled state.\\

Searching numerically for the values of $c$, $d$, $e$ and $f$ which minimize $ C(\rho)$ while imposing $\rho$ to verify the properties of a density matrix, one determines a lower bound for the concurrence. Let us consider the particular case of a maximally entangled pure state of the form 
\begin{equation}
\label{eq:state}
\begin{split} |\Psi (\alpha)\rangle = \frac{1}{\sqrt{2}} & \left( |SS\rangle + e^{2i \alpha} |AA\rangle \right)\\ 
= \frac{1}{\sqrt{2}}  & \Big[  \cos (\alpha) (|LL\rangle + |RR\rangle )  \\
 &  - i \sin (\alpha) (|LR\rangle + |RL\rangle )  \Big] 
\end{split}
\end{equation} 
where $|S\rangle= 1/\sqrt{2} (|L\rangle + |R\rangle)$ is the (symmetric) ground state of the double-well potential and $|A\rangle= 1/\sqrt{2} (|L\rangle - |R\rangle)$ is the (antisymmetric) first excited state. As illustrated in Fig.~\ref{fig:conc} (left), the lower bound on the concurrence depends strongly on $\alpha$. For $\alpha=0$ (state $\rho_a$ on Fig.~\ref{fig:G2e}) and for $\alpha= \pi/2$, one can in principle demonstrate that the state is maximally entangled. Experimentally the contrast of the atomic correlation pattern may be reduced by experimental imperfections, for example the finite resolution of the imaging system. This will reduce the lower bound for the concurrence, as illustrated on Fig.~\ref{fig:conc}. On the contrary, for $\alpha=\pi/4$, corresponding to the state $\rho_c$ on Fig.~\ref{fig:G2e}, the correlation function is identical to the one of the fully mixed state $\rho_d$ and therefore this method does not provide any information on the entanglement in this case. On the other side the method is perfectly suited to characterize a Werner state defined as
\begin{equation}
\label{eq:Werner}
\rho_W= \frac{1}{4} \left( \begin{array}{cccc}
1-W & 0 & 0 & 0 \\
0 & 1+W & -2W & 0 \\
0 & -2W & 1+W & 0 \\
0 & 0 & 0 & 1-W  \end{array} \right),
\end{equation} 
whose degree of mixing depends on $W \in [-1/3,1]$. As illustrated on Fig.~\ref{fig:conc} (right), the exact state concurrence can be determined.


\section{Measurement of a Bell parameter}
\label{sec:Bell}

Since the Bell states $ |\Psi (0)\rangle = 1/\sqrt{2} \left( |LL\rangle +  |RR\rangle \right)$, $ |\Psi (\pi /2)\rangle = 1/\sqrt{2} \left( |LR\rangle +  |RL\rangle \right)$,  $  1/\sqrt{2} \left( |LL\rangle -  |RR\rangle \right)$, and $  1/\sqrt{2} \left( |LR\rangle -  |RL\rangle \right)$ can in principle be fully reconstructed from the correlation measurements, one may wonder how these measurements could allow testing a Bell's inequality. The scheme we described in this article closely relates to the scheme used by Rarity and Tapster to demonstrate a violation of a Bell's inequality in momentum for photons \cite{Rarity1990}, where four momentum modes were mixed two-by-two on beam splitters and the coincidences between the outputs were measured in function of the dephasings $\phi_1$ and $\phi_2$ applied at one input of each beam splitter . Measuring intensity at the two outputs of a beam splitter as a function of an input dephasing $\phi$ is formally equivalent to measuring the far-field intensity resulting from a double-slit experiment as a function of the position (shifted by $\pm 1/4$ of a period, and after correction from the envelop). This is still true when measuring intensity correlations at the output of two independent beam-splitters/double-slit systems, such that we can derive a parameter $E$ from the values of $G^{(2)}(p_1,p_2)$ measured in four points: 
\begin{equation}
\label{eq:E}
E(p_1,p_2)= \frac{g^{(2)}_{--} + g^{(2)}_{++} -  g^{(2)}_{-+} -  g^{(2)}_{+-}} { g^{(2)}_{--} + g^{(2)}_{++} + g^{(2)}_{-+} +  g^{(2)}_{+-} }
\end{equation}
with $g^{(2)}_{\pm \pm}= G^{(2)}(p_1 \pm \pi/4k ,p_2 \pm \pi/4k)/ \left(\mathcal{A}(p_1)^2 \mathcal{A}(p_2)^2 \right)$. From $E(p_1,p_2)$ one obtains the Bell parameter 
$$S_{Bell}= \mathrm{max}  | E(p_1,p_2) - E(p_1,p_2') + E(p_1',p_2) + E(p_1',p_2')| $$
where the maximum is taken over all $\{p_1,p_1',p_2,p_2'\}$ configurations. The remarkable advantage of the double-slit (or double-well) configuration is that for a single experimental configuration one obtains $G^{(2)}(p_1,p_2)$ for a large range of $(p_1,p_2)$, while in the Rarity-Tapster scheme one needs to perform the experiment for different parameter settings $(\phi_1,\phi_2)$. Furthermore, the Bell parameter could be computed when averaging over many more points of $G^{(2)}(p_1,p_2)$, therefore reducing the associated error.\\ 

\begin{figure*}
\centering
\includegraphics[width=14cm]{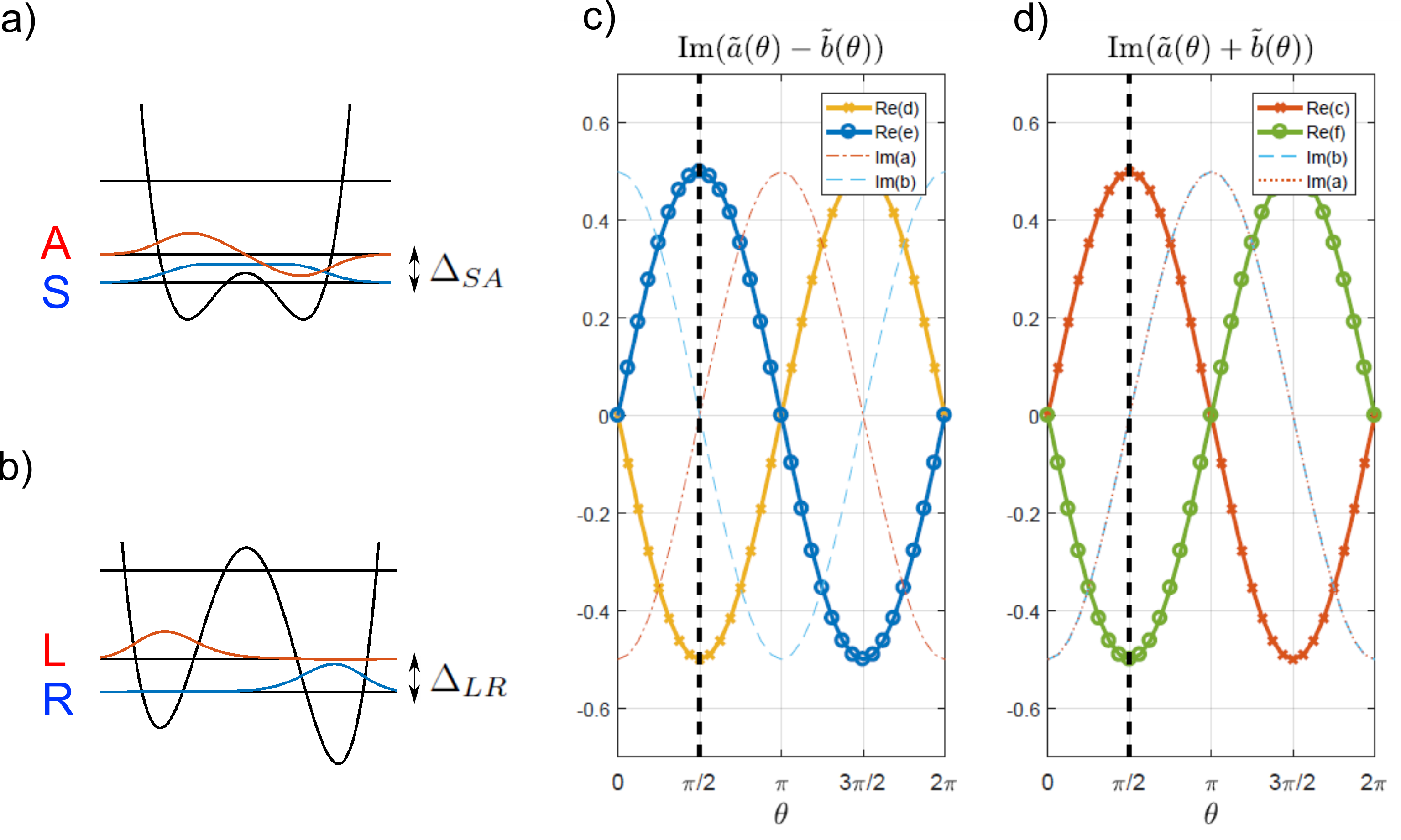}
\caption{\label{fig:tomo} By transforming the double-well potential one can perform a full tomography of the two-atom state. Two different transformations come into play: (a) In coupled double wells the atoms can tunnel and the state undergo rotation in the $\{|L\rangle, |R\rangle \}$ basis [see Eq.~(\ref{eq:rotrho})]. (b) In decoupled, asymmetric double wells  a phase difference accumulates between the left and right wells. 
For a $\pi/2$ phase difference, the imaginary parts of the $c$, $d$, $e$, $f$ elements of the initial density matrix $\rho$ are projected onto the real part of the new density matrix $\rho'$ [see Eq.~(\ref{eq:rot_im})]. 
(c) and (d): After the rotation illustrated in (a), decomposition of the diagonal coherences of the rotated density matrix $\tilde{\rho}$ in function of the elements of the non-rotated density matrix $\rho$. Only the non-zero coefficients of the decomposition are plotted here. 
For $\theta=\pi/2$, $\Im(\tilde{a}- \tilde{b})=1/2 \Re(d-e)$ and $\Im(\tilde{a} + \tilde{b})=1/2 \Re(c-f)$. Since $d+e$ and $c+f$ are known, on can reconstruct the real parts of $c$, $d$, $e$, and $f$. Applying the same operation to $\rho'$, one can reconstruct their imaginary parts.
}
\end{figure*}

We plot on Fig.~\ref{fig:conc} (bottom left) the Bell parameter which could be computed from the correlation measurements for pure states of the form $|\Psi (\alpha) \rangle$, for different contrasts of the correlation pattern. The Bell parameter $S_{Bell}$ is proportional to the contrast, such that observing a violation of a Bell's inequality ($S_{Bell}>2$) for a Bell state (for example $ |\Psi (0)\rangle $ or $ |\Psi (\pi /2)\rangle $) requires a minimum contrast of $1/\sqrt{2}$. The Bell violation is thus much more sensitive on the contrast than the minimum concurrence. While being both entanglement witnesses, the Bell parameter $S_{Bell}$ and the concurrence do not provide the same information \cite{Munro2001}. $S_{Bell}$ only depends on the $a$ and $b$ coefficients describing, for a state $\rho$ of the form (\ref{eq:rho}), the inter-well coherences shared by the two atoms: From the decomposition of Eq.~(\ref{eq:G2comp}) and due to the periodicity of the $\phi_{ijkl}(p_1,p_2)$ functions, Eq.~(\ref{eq:E}) simplifies to $ E(p_1,p_2)= 2 \Re(a) \cos(2k(p_1-p_2)) + 2 \Re(b) \cos(2k(p_1+p_2))$. \\

When measuring $G^{(2)}(p_1,p_2)$ in the far-field of the double-wells one performs thus a Bell test in the $\{|L\rangle, |R\rangle \}$ basis. While $|\Psi (\alpha) \rangle$ is always a maximally entangled state and would maximally violate a Bell inequality in the  $\{|S\rangle, |A\rangle \}$ basis for any $\alpha$, the Bell parameter accessible from the measurements we described depends on $\alpha$ and can only be maximal for the Bell states $ |\Psi (0)\rangle$ and $ |\Psi (\pi /2)\rangle $. These measurements are in contrast perfectly suited for the Werner state $\rho_W$ (Fig.~\ref{fig:conc}, bottom right), since its entanglement is described by the $a$ coefficient only. 
The $G^{(2)}(p_1,p_2)$ measurement described in this article is thus a powerful tool to detect Bell correlations in the $\{|L\rangle, |R\rangle \}$ basis.


\section{Full tomography method}
\label{sec:tom}

In order to perform a full tomography, one still needs to determine independently the $c$, $d$, $e$ and $f$ components of Eq.~(\ref{eq:rho}). We therefore would like to transform the state $\rho$ into a state $\tilde{\rho}$ from which the initial values of  $c$, $d$, $e$ and $f$ can be traced back. The real and imaginary parts of these parameters are accessible with two different transformations, which are easily performed in double-well potentials as depicted on Figs.~\ref{fig:tomo}.a and \ref{fig:tomo}.b.\\ 

For the real parts it is sufficient to perform a rotation in the $\{|L\rangle, |R\rangle \}$ basis (Fig.~\ref{fig:tomo}.a): When the tunneling between the two wells is switched on, i.e. for a non-negligible energy difference $\Delta_{SA}$ between the lowest two eigenstates $|S\rangle$ and $|A\rangle$ of the double-well potential, a phase difference $\theta$ accumulates between these two states. This gives rise to bosonic Josephson dynamics \cite{Albiez2005}, which can be used to probe the coherences between the wells \cite{Kessler2014}. Each atom undergoes a rotation
\begin{equation}
R_i(\theta) = \left( \begin{array}{cc}
\cos \theta/2 & -i \sin \theta/2 \\
-i \sin \theta/2 &  \cos \theta/2 \end{array} \right)
\end{equation}
such that the final density matrix is 
\begin{equation}
\label{eq:rotrho}
\tilde{\rho}(\theta) = \left( R_1(\theta) \otimes  R_2(\theta) \right) \rho  \left( R_1(\theta)^{-1} \otimes  R_2(\theta)^{-1} \right)
\end{equation} 
This rotated state can be characterized as described in section~\ref{sec:G2} and the coherences $\tilde{a}(\theta)$ and $\tilde{b}(\theta)$ extracted. We show in Fig.~\ref{fig:tomo}.c and \ref{fig:tomo}.d how the real parts of  $c$, $d$, $e$ and $f$ are deduced when measuring $\tilde{a}(\pi/2)$ and $\tilde{b}(\pi/2)$.\\ 

For reconstructing the imaginary parts of  $c$, $d$, $e$ and $f$ we introduce another transformation of the double-well potential: With an asymmetry between the left and right well (energy difference $\Delta_{LR}$ on Fig.~\ref{fig:tomo}.b) a phase difference $\phi$ between these wells accumulates linearly with time \cite{Berrada2013}, transforming the state $\rho$ into
\begin{equation}
\label{eq:rot_im}
\rho'= \left( \begin{array}{cccc}
p_{LL} & c \; e^{i \phi} &  d \; e^{i \phi} & b \; e^{2 i \phi}\\
c^* \; e^{- i \phi} & p_{LR} & a & e \; e^{i \phi} \\
d^*\; e^{-i \phi} & a^* & p_{RL} & f \; e^{i \phi}\\
b^* \; e^{-2 i \phi} & e^* \; e^{-i \phi}&  f^* \; e^{-i \phi}& p_{RR}  \end{array} \right)
\end{equation}
When switching the double-well potential back to the initial symmetric configuration after a delay corresponding to $\phi=\pi/2$, one gets $\Re(c')=- \Im(c)$ (and similarly for $d$, $e$ and $f$) and can measure these elements when performing the rotation in the $\{|L\rangle, |R\rangle \}$ basis of Eq.~(\ref{eq:rotrho}) illustrated on Fig.~\ref{fig:tomo}.a.\\

Thus, thanks to the tunability of double-well potentials, the full density matrix can in principle be reconstructed.

\section{Proposed experimental implementation}
\label{sec:xp}

\begin{figure}[b]
\centering
\def\svgwidth{\columnwidth}
\includegraphics[width=\columnwidth]{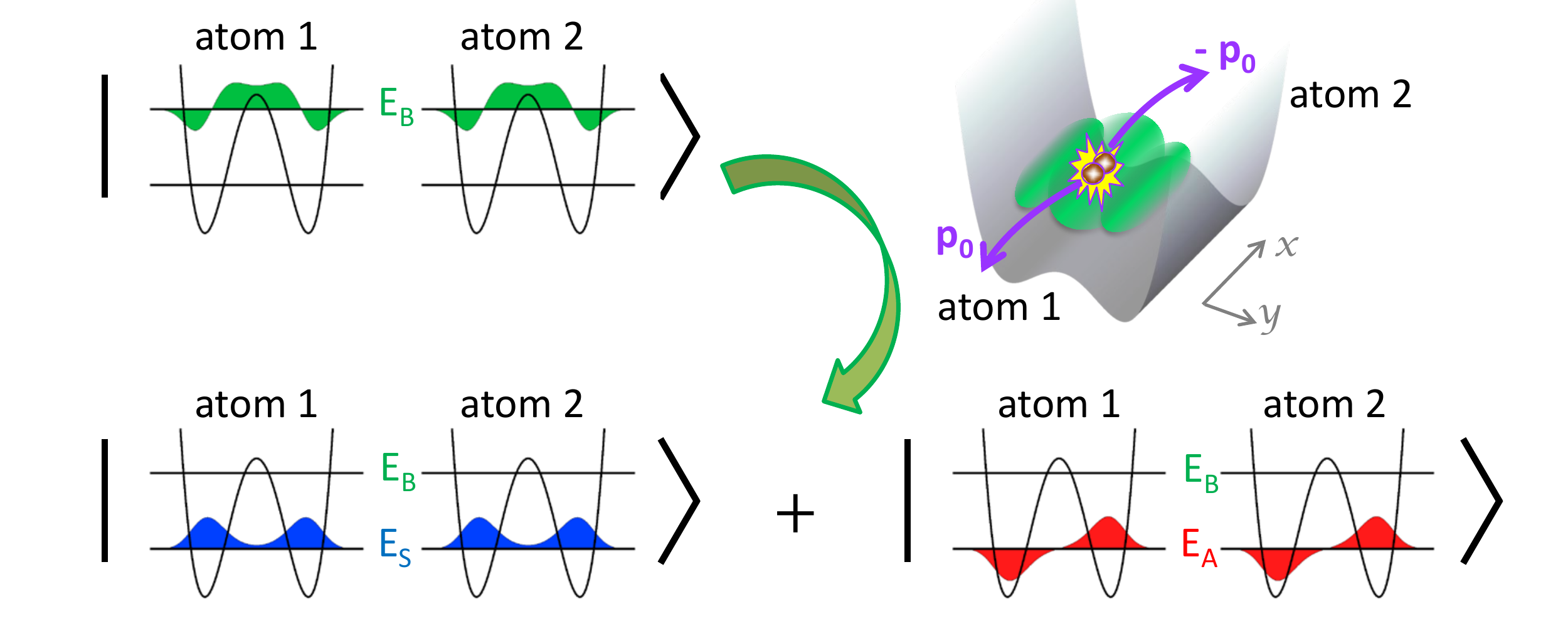}
\caption{\label{fig:WF} An entangled two-atom state in double wells (bottom) is obtained for example through four-wave mixing, starting from a BEC in an excited transverse state of the potential (top left). The two atoms have opposite momenta $\pm p_0$ along the longitudinal axis $x$ (top right).}
\end{figure}

Pairs of entangled atoms in a double-well potential can be obtained in a number of different ways. We envision here an experiment where each pair is the output of an atomic four-wave mixing process. The source is a Bose-Einstein condensate trapped in an elongated potential which has a double-well shape along one of the strongly confining axes ($y$), as illustrated on Fig.~\ref{fig:WF}. One triggers the four-wave mixing process by transferring the BEC to the second excited state $|B\rangle$ of the transverse double-well potential (in green on Fig.~\ref{fig:WF}) \footnote{This motional state preparation can be performed by modulating the trapping potential. The shape of the modulation is defined numerically by optimal control of the spatial wave function. A similar state preparation applied to a single-well potential was described in \cite{Frank2016}.}.
We note this state $|p_{\parallel}=0,B\rangle$, where $p_{\parallel}$ is the momentum along the weakly confining longitudinal axis $x$. This state provides two degenerate input modes for the four-wave mixing. Energy and momentum conservation select two different pairs of output modes: $|p_{\parallel}=\pm p_S,S\rangle$ where two atoms are emitted to the transverse ground state $|S\rangle$ with opposite longitudinal momenta $p_S= \sqrt{2m (E_B-E_S)}$ ($m$ being the atom mass and $E_i$ the energy of the eigenstate $|i\rangle$ of the double-well potential ) and $|p_{\parallel}=\pm p_A,A\rangle$ where two atoms are emitted to the transverse first excited state $|A\rangle$ with opposite longitudinal momenta $p_A= \sqrt{2m (E_B-E_A)}$. For negligible tunnel coupling, $E_A=E_S$ such that the two pairs of output modes overlap in momentum: ${p_S = p_A = p_0}$. 
We define the first (second) atom as the one with longitudinal momentum $p_0$ ($-p_0$). 
The expected output state of the four-wave mixing is therefore the Bell state
\begin{equation}
\label{eq:stateFWF}
\begin{split} |\Psi_{out}\rangle &= 1/\sqrt{2} \left(|SS\rangle + |AA\rangle \right) = |\Psi (\alpha=0)\rangle\\ 
&= \frac{1}{\sqrt{2}}  (|LL\rangle + |RR\rangle ) 
\end{split}
\end{equation} 
From this state, any state $|\Psi (\alpha)\rangle$ (Eq.~(\ref{eq:state})) could be prepared by switching on the tunnel coupling (\textsl{i.e.} the energy difference $\Delta_{SA}$ between the  $|S\rangle$ and  $|A\rangle$ eigenstate), as illustrated on Fig.~\ref{fig:tomo}.a and described in section \ref{sec:tom}.  \\

For characterizing the state we need to measure coincidences in the momentum distribution and therefore a single-atom resolution is required. Spatially resolved single-atom detectors are based, for example, on fluorescence imaging combined with an electron multiplying charge coupled device (EMCCD) camera \cite{Bucker2009,Bergschneider2018} or on multichannel plates \cite{Schellekens2005}. The spatial resolution of the imaging system is a key parameter: It could reduce the contrast of the correlation pattern fringes and prevent detecting a Bell's inequality violation. The atom momentum distribution is mapped on the position distribution at the detector after a large time-of-flight propagation. Our characterization method requires identifying single pairs: When working with low atom numbers source BECs and short pair production times, one reaches a regime where single-pair emission is frequent. Post-selecting such events requires a high detection efficiency, as provided by fluorescence imaging.

\section{Conclusion}

We have described in this article various methods for characterizing two-atom states in double-well potentials. In contrast to what was done for photons based on analogous measurements, the only hypothesis we use here is that only the first two single-particle eigenstates of the double-well potential are accessible. We have explained how, without modifying the shape of the potential, the two-particle density matrix can be partially reconstructed from the two-particle correlations in position and momentum space and a lower bound for entanglement determined. This method is particularly interesting for strongly entangled states, since from the same measurement a Bell's inequality violation can be probed. We note that it also applies to analogous systems where each atom can occupy two modes which are separated in the near field and can interfere in the farfield. We finally described how, using standard manipulations in the double-well potential, the full density matrix can be reconstructed.

\section{Acknowledgements}
We thank R. B{\"u}cker for inspiring this work. MB was supported by the EU through the Marie Sk\l{}odowska Curie Grant ETAB (MSCA-IF-2014-EF 656530) and by the Austrian Science Fund (FWF) through the Lise Meitner Grant CoPaNeq (M2088-M27). JS acknowledges support by the ERC Advanced Grant QuantumRelax. 



\end{document}